\documentclass[a4paper,12pt]{article}
\usepackage{amsfonts}
\usepackage{latexsym}
\usepackage{amsmath}
\usepackage{amssymb}
\usepackage{amssymb}
\usepackage{slashed}
\usepackage{upgreek}
\usepackage{mathrsfs}
\usepackage{bbm}

\usepackage{amsthm}

\theoremstyle{remark}

\usepackage[utf8]{inputenc}
\usepackage[english]{babel}

\theoremstyle{definition}

\allowdisplaybreaks

\usepackage{relsize}
\usepackage{graphicx}
\usepackage{comment}

\renewcommand{\thefootnote}{\fnsymbol{footnote}}

\def\appendix#1{\addtocounter{section}{1}\setcounter{equation}{0}
\renewcommand{\thesection}{\Alph{section}}
\section*{Appendix \thesection\protect\indent \parbox[t]{11.15cm}{#1}}
\addcontentsline{toc}{section}{Appendix \thesection\ \ \ #1}}

\font\mybb=msbm10 at 11pt

\def\bb#1{\hbox{\mybb#1}}

\def\bR {\bb{R}}

\newcommand{\bea}{\begin{eqnarray}}
\newcommand{\eea}{\end{eqnarray}}

\usepackage[usenames,dvipsnames,svgnames,table]{xcolor}	% AH -- REMOVE BEFORE UPLOAD
\usepackage[normalem]{ulem}				% AH -- REMOVE BEFORE UPLOAD
\usepackage{microtype}
\usepackage[colorlinks, citecolor=blue, linkcolor=blue, urlcolor=Maroon, filecolor=Maroon, linktocpage=true]{hyperref} % AH

\usepackage{chngcntr}
\counterwithout{equation}{section}

\begin{document}

\begin{center}
%\today
\vspace*{-1.0cm}
\begin{flushright}
%\normalsize{\texttt{ZMP-HH/17-24}}\\
\end{flushright}
%\hfill hep-th/yymmnnn \\
%\hfill UB-ECM-PF-06-43 \\
%\hfill DMUS--MP--13/06 \\

%\vspace{2.0cm} {\Large \bf Covariant form hierarchies, integrability  and Killing spinor equations } \\[.2cm]

\vspace{2.0cm} {\Large \bf Symmetries, Spinning Particles and the  TCFH of D=4,5 Minimal Supergravities } \\[.2cm]

\vskip 2cm
G.\,  Papadopoulos and  E.\, P\'erez-Bola\~nos
\\
\vskip .6cm

%\begin{small}
%${}^1$ \textit{Department of Mathematics, King's College London
%\\
%Strand, London WC2R 2LS, UK}\\
%\texttt{sebastian.lautz@kcl.ac.uk}\\
%\end{small}
%\vskip0.5cm

\begin{small}
\textit{Department of Mathematics
\\
King's College London
\\
Strand
\\
 London WC2R 2LS, UK}
 \\*[.3cm]
 \texttt{george.papadopoulos@kcl.ac.uk}
 \\
\texttt{edgar.perez$\underline{~}$bolanos@kcl.ac.uk}
\end{small}
\\*[.6cm]

\end{center}

\vskip 2.5 cm

\begin{abstract}
\noindent
We find that spinning particles  with suitable couplings propagating in certain supersymmetric  backgrounds of   $D=4$, $N=2$ and $D=5$, $N=1$ minimal supergravities are invariant under symmetries generated by  the twisted covariant form hierarchies  of these  theories. We also compare our results with the symmetries  of spinning particles  generated by  Killing-Yano forms which are responsible for the separability properties of some gravitational backgrounds.

\end{abstract}

%\vskip 1cm

%{\small Keywords: spacetime geometry; black holes; special Lorentzian structures; G-structures}

\newpage

\renewcommand{\thefootnote}{\arabic{footnote}}

\section*{}

It has been known for sometime that the geodesic equation as well as several classical field equations, like Klein-Gordon, Hamilton-Jacobi, Dirac and Maxwell equations,  in some gravitational backgrounds can be separated, for reviews see \cite{revky, frolov} and references within. A celebrated example of such background is the Kerr black hole solution \cite{carter-a, carter-b, carter-c, chandrasekhar}. The separability properties of gravitational backgrounds are due to the existence of hidden symmetries generated by  rank 2 Killing-St\"ackel tensors\footnote{This is a symmetric tensor $K$ which satisfies $\nabla_\mu K_{\nu\rho}=\nabla_{(\mu} K_{\nu\rho)}$.} and the observation that the Killing tensors are  squares of Killing-Yano (KY) forms \cite{penrose-a, penrose-b, floyd}, see footnote 2 for the definitions of these forms. Furthermore it was pointed out in \cite{gibbons} that the KY forms generate fermionic symmetries, like supersymmetries, in actions that model the propagation of  spinning particles \cite{berezin, howe-a, howe-b, barducci}  in such backgrounds. As a result, there is a relation between the separability properties of gravitational  backgrounds and conservation laws in spinning particle systems.

Recently, it has been shown in \cite{gp} that the conditions satisfied by the Killing spinor form bi-linears of supersymmetric backgrounds in all supergravity theories  including higher curvature corrections can be arranged as twisted covariant form hierarchies (TCFH) \cite{jggp}. This means that there is a connection $\nabla^{\cal F}$ which depends on the fluxes ${\cal F}$ of a theory such that
\bea
\nabla_X^{\cal F}{\cal X}=i_X{\cal P}+X\wedge {\cal Q}~,
\eea
where ${\cal X}$ is a multi-form spanned by all the (Killing spinor) form bilinears,  ${\cal P}$ and ${\cal Q}$ are multi-forms which depend on ${\cal F}$ and the form bilinears and $X$ is a vector field on the spacetime. We have also denoted with $X$ the associated one-form to  $X$, $X(\cdot)=g(\cdot, X)$. An alternative way to state the above condition is  that the highest weight representation in the decomposition of the tensor $\nabla^{\cal F}{\cal X}$ in orthogonal irreducible representations vanishes. In general the connection $\nabla^{\cal F}$ is not form degree preserving.

 A consequence of the TCFH is that ${\cal X}$ satisfies a generalisation\footnote{For the standard CKY equation $\nabla^{\cal F}=\nabla$, where $\nabla$ is the Levi-Civita connection. A form $\chi$ that satisfies the CKY equation is KY provided it is also co-closed $\delta\chi=0$. While $\chi$ is a closed conformal KY  (CCKY) form, iff $d\chi=0$.}  of the Conformal-Killing-Yano (CKY) equation with respect to the connection
 $\nabla^{\cal F}$, i.e. form bilinears satisfy the condition
 \bea
(\nabla_X^{\cal F}{\cal X})\vert_p=i_X (d^{\cal F}{\cal X}\vert_{p+1})-{1\over D-p+1} X\wedge (\delta^{\cal F}{\cal X})\vert_{p-1}~,
\label{tcfh20}
\eea
where we have restricted the equation to a form bilinear of degree $p$ and $D$ is the spacetime dimension.   The operations $d^{\cal F}$ and $\delta^{\cal F}$ are defined by the above equation and denote exterior differentiation and its (formal) adjoint with respect to the connection $\nabla^{\cal F}$. In particular, $d^{\cal F}$ is defined by skew-symmetrising the indices in (\ref{tcfh20}) while  $\delta^{\cal F}$ is defined by taking a contraction of (\ref{tcfh20}) with respect to the spacetime metric.

 The validity  of (\ref{tcfh20}) for all supersymmetric solutions raises the question on  whether the form bilinears can be used to investigate the separability properties of these backgrounds, and on  whether they  generate symmetries in spinning  particles  propagating on such backgrounds.
In this article,  we shall demonstrate that the  form bilinears of a large class of supersymmetric $D=4$, $N=2$ and $D=5$, $N=1$ minimal supergravity backgrounds generate symmetries in spinning particle actions with appropriate couplings.  The key observation is that some of the conditions for invariance of the particle actions of \cite{rcgp} under some fermionic  transformations can also be expressed as  TCFHs.  In this case, the associated TCFH connection  depends of the couplings of the particle action and  acts on forms that determine the infinitesimal  fermionic symmetries of the system. Thus the task is to match the  TCFHs of  supersymmetric backgrounds with those of the spinning particle symmetries after an appropriate identification of the supergravity fields with the couplings of the particle system and of the form bilinears with the forms that generate the fermionic symmetries, respectively. We shall demonstrate that this can be achieved in a variety of cases.
We shall also comment on the use of the form bilinears to investigate the separability properties of supersymmetric backgrounds.

%\section{Minimal D=4, N=2 supergravity}
The supercovariant connection of minimal $D=4$, $N=2$ supergravity  is
 \bea
 {\cal D}_\mu\equiv \nabla_\mu+{i\over 4}F_{ab}\Gamma^{ab}\Gamma_\mu~,
 \eea
 where $F$ is a 2-form field strength, $dF=0$. The field equations also imply that $F$ is co-closed, $d{}^*F=0$. If $\epsilon$ is a Killing spinor, ${\cal D}_\mu\epsilon=0$, the form bilinears of the theory up to a Hodge duality are
\bea
 &&f=\langle \epsilon, \epsilon\rangle_D~,~~~h=\langle \epsilon,\Gamma_5 \epsilon\rangle_D~,~~~K=\langle \epsilon, \Gamma_a\epsilon\rangle_D\, e^a~,~~~
 \cr
 &&
 Y^1=\langle \epsilon, \Gamma_a\Gamma_5\epsilon\rangle_D\, e^a~,~~~ Y^3+iY^2=\langle\tilde{\epsilon},\Gamma_a\Gamma_5\epsilon\rangle_D\, e^a~,~~~
 \cr
 &&
\omega^1={1\over2}\langle \epsilon, \Gamma_{ab}\epsilon\rangle_D\, e^a\wedge e^b~,~~~
 \omega^3+i\omega^2={1\over 2}\langle\tilde{\epsilon}, \Gamma_{ab} \epsilon\rangle_D\, e^a \wedge e^b ~,~~~
\eea
where the spacetime metric $g=\eta_{ab} e^a e^b$ with $e^a=e^a_\mu dx^\mu$ a local co-frame,  $\langle\cdot,\cdot\rangle_D$ is the Dirac inner product, $C$ is a charge conjugation matrix such that $C*\Gamma_a=-\Gamma_a C*$ and $C*C*=-1$,  and $\tilde{\epsilon}=C*\epsilon$. $C=\Gamma_3$ in the conventions of \cite{review}.  Observe that if $\epsilon$ is a Killing spinor so is $\tilde\epsilon$. The TCFH of the theory \cite{gp} reads
\bea
&&
\nabla_\mu f=i F_{\mu \nu}  K^\nu~,~~~
\nabla_\mu h= {}^{*}F_{\mu \nu}K^\nu~,~~~
\nabla_\mu K_\nu=ifF_{\mu\nu}-h\,{}^*F_{\mu\nu}~,~~~
\cr
&&
\nabla_\mu Y^r_\nu+2{}^*F_{\mu \rho}\omega^{r\rho}{}_{\nu}=2{}^*F_{[\mu|\rho|}\omega^{r\rho}{}_{\nu]}-{1\over 2}g_{\mu\nu}{}^*F_{\rho\lambda}\omega^{r\rho\lambda}~,~~~r=1,2,3~,
\cr
&&
\nabla_\mu \omega^r_{\nu\rho}-4\,{}^*F_{\mu[\nu}Y^r_{\rho]}=-3\,{}^*F_{[\mu\nu}Y^r_{\rho]}-2\,g_{\mu[\nu}{}^*F_{\rho]\lambda}Y^{r\lambda}~,~~~r=1,2,3~.
\label{tcfh1}
\eea
In what follows we shall also consider the TCFH associated with the dual 2-forms $\chi^r$ of $\omega^r$ which can be defined as
\bea
\chi^1=-{i\over2}\langle \epsilon, \Gamma_{ab}\Gamma_5\epsilon\rangle_D\, e^a\wedge e^b~,~~~
 \chi^3+i\chi^2=-{i\over 2}\langle\tilde{\epsilon}, \Gamma_{ab}\Gamma_5 \epsilon\rangle_D\, e^a \wedge e^b ~.~~~
\eea
One can show that the Killing spinor equations imply the TCFH
\bea
&&\nabla_\mu Y^r_\nu+2{}F_{\mu \rho}\chi^{r\rho}{}_{\nu}=2{}F_{[\mu|\rho|}\chi^{r\rho}{}_{\nu]}-{1\over 2}g_{\mu\nu}{}F_{\rho\lambda}\chi^{r\rho\lambda}~,~~~r=1,2,3
\cr
&&
\nabla_\mu \chi^r_{\nu\rho}-4\,F_{\mu[\nu}Y^r_{\rho]}=-3\,F_{[\mu\nu}Y^r_{\rho]}-2\,g_{\mu[\nu}F_{\rho]\lambda}Y^{r\lambda}~,~~~r=1,2,3~.
\label{tcfh1a}
\eea
It is clear from (\ref{tcfh1}) that $K$ is a Killing vector which also leaves $F$ invariant, ${\cal L}_KF=0$.

To determine whether the above TCFH generates symmetries in a particle system propagating in the supersymmetric backgrounds of $D=4$, $N=2$ supergravity consider  the worldline action
\bea
S=\int dt d\theta \big(-{i\over2} g_{\mu\nu} D\phi^\mu \partial_t \phi^\nu+i q_{\mu\nu} D\phi^\mu D\phi^\nu \psi+ {1\over2} \psi D\psi\big)~,
\label{paction1}
\eea
 where $\phi$ is a bosonic and $\psi$ is a fermionic superfield that depend on the worldline time $t$ and the odd coordinate $\theta$ and $D=\partial_\theta+i\theta \partial_t$ with $D^2=i\partial_t$. The fields have components $\phi=\phi\vert, \lambda=D\phi\vert$, $\psi=\psi\vert$ and $A=D\psi\vert$, where the restriction means evaluation at $\theta=0$.  $\phi$ and $A$ are worldline bosons while the rest of the components are worldline fermions. The couplings of the theory are the spacetime metric $g$ and the 2-form $q$  which depend of $\phi$. Later $q$   will be identified with either $F$ or its dual ${}^*F$. This action is manifestly invariant under one worldline supersymmetry.  To write the action above we adopted the reality conditions
\bea
(i\partial_t)^*=i\partial_t~,~~~\theta^*=\theta~,~~~\phi^*=\phi~,~~~\phi^*=\phi~,~~~\psi^*=-\psi~,~~(\chi \lambda)^*=\chi^* \lambda^*~,
\eea
for every two worldline fermions $\chi$ and $\lambda$. With these reality conditions the coupling of the theory  $g$ and  $q$ are real. Such a choice of reality conditions is not unique.  For example one could have chosen  $\psi^*=\psi$ at the cost of removing the imaginary unit $i$ from the coupling term $q(D\Phi)^2\psi$ in the action. However such a choice is not suitable for the application we are investigating. The action (\ref{paction1}) is a special case of a general class of actions for spinning particles  presented in \cite{rcgp}.

Identifying $q$ with either $F$ or ${}^*F$, the  Killing vector $K$ of the TCFH (\ref{tcfh1})  generates   the infinitesimal transformation
\bea
\delta\phi^\mu= a K^\mu~,~~~\delta\psi=0~,
\eea
which is a symmetry of the action.  Thus the isometries of the supersymmetric backgrounds of  $D=4$, $N=2$ supergravity generate a symmetry in the particle system  action (\ref{paction1}).

It remains to see whether the remaining conditions of the TCFH (\ref{tcfh1}) are associated with symmetries.  For this  consider the fermionic transformations
\bea
\delta\phi^\mu=i\alpha I^\mu{}_\nu D\phi^\nu+ \alpha L^\mu \psi~,~~~\delta\psi=i \alpha M_\mu \partial_t \phi^\mu~,
\label{sym1}
\eea
where $I, L$ and $M$ depend on $\phi$ and $\alpha$ is a anti-commuting infinitesimal parameter. The reality condition for $\alpha$ is chosen as $\alpha^*=-\alpha$ which has as a consequence the presence of imaginary unit in the $ID\phi$ term of the infinitesimal transformation of $\phi$. Again this is essential for the application we shall present below. With this choice of reality condition the tensors $I$, $L$ and $M$ are real. After some simplification the conditions for the invariance of the  action (\ref{paction1}) under the   infinitesimal transformations (\ref{sym1})   can be expressed\footnote{The are many  inequivalent ways to write the conditions for the invariance of the action (\ref{paction1}) under the transformations (\ref{sym1}). However, the form given below is suitable for the investigation of this example.}  as
\bea
&&\nabla_\mu I_{\nu\rho}-4 q_{\mu[\nu} M_{\rho]}=-6 q_{[\mu\nu} M_{\rho]}~,~~~I_{[\nu\rho]}=I_{\nu\rho}~,
\cr
&&L_\mu=-M_\mu~,~~~\nabla_\mu M_\nu+2 q_{\mu \rho} I^\rho{}_\nu=0~,~~~
dq_{\lambda[\mu\nu} I^\lambda{}_{\rho]}=0~.
\label{tcfhs1}
\eea
Note that if instead we had chosen as reality conditions $\psi^*=\psi$ and $\alpha^*=\alpha$ with the rest remaining the same, the sign of the term $qI$ in the conditions above would have been different.  The differential conditions as stated in (\ref{tcfhs1}) on the tensors associated to the infinitesimal transformations (\ref{sym1}) are in a TCFH form with connection which depends  on the coupling $q$ of the theory.

To compare (\ref{tcfh1}) with (\ref{tcfhs1}), one has to consider three copies of the transformation (\ref{sym1}) generated by the tensors $I^r$ and $M^r$, $r=1,2,3$ and set
\bea
I^r=\omega^r~,~~~M^r=Y^r~,~~~q={}^*F~.
\label{idx}
\eea
With these identifications the connection part of the TCFHs in (\ref{tcfh1}) and (\ref{tcfhs1}) match. However consistency requires that
the right-hand-side of the last two equations in (\ref{tcfh1}) must vanish.
As a result, $\omega^r, Y^r$ are parallel with respect to the TCFH connection.  Note also that $d{}^*F=0$ as a consequence of the field equations and so the last condition in (\ref{tcfhs1}) is satisfied.

The commutators of the symmetries (\ref{sym1}) can be easily be computed \cite{rcgp}. After the identification (\ref{idx}) and for the backgrounds investigated below, it can be easily seen that that they do not close to the standard supersymmetry algebra $\{Q^r, Q^s\}=\delta^{rs} H$ in one dimension.  This is in agreement with the commutators of  the fermionic symmetries generated by KY forms in \cite{gibbons}.

The supersymmetric solutions of minimal $D=4$, $N=2$ supergravity have been classified in \cite{tod}. A class of backgrounds for which the right-hand-side of the last two equations in (\ref{tcfh1})  vanishes are those that admit a null Killing spinor, i.e. a spinor for which the bilinear $K$ is null. For all such backgrounds, one can demonstrate as a consequence of the Killing spinor equations that the non-vanishing components of the fluxes and form bilinears are
\bea
&&K=K_- e^-~,~~~Y^r=Y_-^r\, e^-~,~~~\omega^r=\omega_{-i}^r\, e^-\wedge e^i~,~~~
\cr
&&F=F_{-i}\, e^-\wedge e^i~,~~~{}^*F={}^*F_{-i}\, e^-\wedge e^i~,~~~
\eea
see \cite{review} for more details, where $(e^+, e^-, e^i)$ is a co-frame such that the metric $g=2 e^+ e^-+\delta_{ij} e^i e^j$, $i,j=1,2$,
i.e. the form bilinears and the flux $F$ are null forms. Using this, one can easily verify that the right-hand-side of the last two equations in (\ref{tcfh1})  vanishes.
Therefore particles systems described by (\ref{paction1}) propagating on backgrounds  with a null Killing spinor and  couplings the spacetime metric $g$ and $q={}^*F$ admit symmetries (\ref{sym1}) generated by the associated form bilinears. Such solutions include for example pp-wave type of backgrounds.

One can also consider the symmetries generated by the TCFH (\ref{tcfh1a}).  The investigation for this is similar to the one we have presented above for the TCFH (\ref{tcfh1}).  The only difference is that in this case $I^r=\chi^r$ and $q=F$.  Thus again the spinning particles  described by the action (\ref{paction1}) with couplings the spacetime metric $g$  and $q=F$ admit symmetries (\ref{sym1}) generated by the form bilinears  $Y^r$ and $\chi^r$.
 
 A similar analysis can be performed for supersymmetric backgrounds  with a time-like Killing spinor, i.e. $K$ is a time-like vector.  However in this case one can show that either the condition ${}^*F_{\mu\nu} Y^{r\nu}=0$  which arises from the comparison of  (\ref{tcfh1}) with (\ref{tcfhs1}) or $F_{\mu\nu} Y^{r\nu}=0$ which arises from the comparison of (\ref{tcfh1a}) with (\ref{tcfhs1}), for all $r=1,2,3$, require    that $F=0$. This is because the 1-forms $Y^r$ are spacelike and span the three spatial directions of the spacetime, see \cite{review}.  The only  solutions with $F=0$  are locally isometric to Minkowski spacetime.

Next let us turn to investigate the TCFH of  $D=5$, $N=1$ minimal supergravity.
The supercovariant connection of the theory is
 \bea
 {\cal D}_\mu\equiv \nabla_\mu-{i\over 4\sqrt3}\big(\Gamma_\mu{}^{\nu\rho}F_{\nu\rho}-4 F_{\mu\nu} \Gamma^\nu\big)~.
 \eea
If $\epsilon$ is a Killing spinor, ${\cal D}_\mu\epsilon=0$  the independent (Killing spinor) form bi-linears up to a Hodge duality operation are
  \bea
 &&f=\langle \epsilon, \epsilon\rangle_D~,~~~K=\langle \epsilon, \Gamma_a\epsilon\rangle_D\, e^a~,~~~\omega^1={1\over2}\langle \epsilon, \Gamma_{ab}\epsilon\rangle_D\, e^a\wedge e^b~,
 \cr
 &&
 \omega^2+i\omega^3={1\over 2}\langle\epsilon, \Gamma_{ab}\tilde{\epsilon}\rangle_D \, e^a\wedge e^b~,~~~
 \eea
 where $e^a$, $a=0,1,2,3,4$ is a co-frame such that the metric is $g=\eta_{ab} e^a e^b$,  $\tilde{\epsilon}=\Gamma_{12}*\epsilon$ in the conventions of \cite{review}.
 If $\epsilon$ is a Killing spinor,  then $i \epsilon$,  $\tilde\epsilon$ and $i\tilde\epsilon$ are also Killing spinors. Supersymmetric backgrounds of this theory preserve either 4 or 8 supersymmetries and have been classified in \cite{gutowski}.
 The conditions imposed by the Killing spinor equation on the form bilinears have been derived in \cite{gutowski}. Writing them   in a  TCFH form, one finds  \cite{gp}  that
\bea
 &&\nabla_\mu f=-{2i\over\sqrt3} F_{\mu\nu} K^\nu~,~~~
 \nabla_\mu K_\nu={1\over \sqrt3}{}^*F_{\mu\nu\rho} K^\rho-{2i\over\sqrt3} F_{\mu\nu}f~,~~~
 \cr
 &&
\nabla_\mu\omega^r_{\nu\rho}-2\sqrt3\, {}^*F_{\lambda\mu[\nu} \omega^{r\lambda}{}_{\rho]}
 =-2\sqrt3\, {}^*F_{\lambda[\nu\rho} \omega^{r\lambda}{}_{\mu]}+ {2\over\sqrt3} g_{\mu[\nu}\, {}^*F_{\rho]\alpha\beta} \omega^{r\alpha\beta}~,
 %\cr
 %&&
 %\nabla_\mu\xi_{NR}-2\sqrt3\, {}^*F_{EM[N} \xi^E{}_{R]}
 %=-2\sqrt3\, {}^*F_{E[NR} \xi^E{}_{M]}+ {2\over\sqrt3} g_{M[N}\, {}^*F_{R]EF} \xi^{EF}~.
 \label{tcfh2}
 \eea
where $\mu,\nu, \rho=0,1,2,3,4$ are spacetime indices and $r=1,2,3$. In what follows, it is also useful to state the TCFH for the form bilinears
\bea
\lambda^1={1\over3!}\langle \epsilon, \Gamma_{abc}\epsilon\rangle_D\, e^a\wedge\cdots\wedge e^c~,
 \lambda^2+i\lambda^3={1\over 3!}\langle\epsilon, \Gamma_{abc}\tilde{\epsilon}\rangle_D \, e^a\wedge \cdots \wedge e^c~,~~~
 \label{lform}
\eea
which are Hodge duals to $\omega^r$. This reads 
\bea
 \nabla_\mu \lambda^r_{\nu_1\nu_2\nu_3}-3\sqrt3 \,{}^*F_{\alpha\mu[\nu_1}\lambda^{r\alpha}{}_{\nu_2\nu_3]}
 =-4\sqrt3 \,{}^*F_{\alpha[\mu\nu_1} \lambda^{r\alpha}{}_{\nu_2\nu_3]}+ 2\sqrt3 g_{\mu[\nu_1} \,{}^*F_{\nu_2|\alpha\beta|}\lambda^{r\alpha\beta}{}_{\nu_3]}~.
 \label{ltcfh}
\eea

 To find whether the above TCFHs are associated with symmetries of a particle system propagating on the spacetime consider the action
 \bea
 S=-{1\over2}\int dt d\theta \big(i\, g_{\mu\nu} D\phi^\mu \partial_t \phi^\nu+{1\over6} c_{\mu\nu\rho} D\phi^\mu D\phi^\nu D\phi^\rho\big)~,
 \label{mact2}
 \eea
 where the superfields $\phi$ are as in (\ref{paction1}) and $c$ is a spacetime 3-form which depends on $\phi$. Actions with such couplings have been considered before in \cite{rcgp}.  This action is manifestly invariant under one supersymmetry.

 Next consider the fermionic symmetry
 \bea
 \delta\phi^\mu=\alpha I^\mu{}_\nu D\phi^\nu~,
 \label{itran2}
 \eea
 where the infinitesimal parameter $\alpha$ satisfies the reality condition $\alpha^*=\alpha$.  Invariance of the action under this fermionic symmetry implies \cite{gpgg} that
 \bea
 &&\hat\nabla_\mu I_{\nu\rho}=\hat\nabla_{[\mu} I_{\nu\rho]}~,~~~I_{\mu\nu}=I_{[\mu\nu]}~,
 \cr
 &&di_Ic-3 i_I dc=0~,
 \label{psusy2}
 \eea
 where
\bea
 \hat\nabla_\mu X^\nu=\nabla_\mu X^\nu+{1\over2} c^\nu{}_{\mu\rho} X^\rho~,
 \label{skewcon}
 \eea
 is a connection with skew-symmetric torsion $c$ and $i_I$ denotes the inner derivation\footnote{The inner derivation of a $n$-form $\chi$ with respect to the vector $(k-1)$-form $L$ is $i_L\chi={1\over (k-1)! (n-1)!} L^\nu{}_{\mu_1\dots \mu_{k-1}} \chi_{\nu \mu_k\dots \mu_{n+k-2}} dx^{\mu_1}\wedge\dots \wedge dx^{\mu_{k+n-2}}$.}  with respect to $I$.

 One can also consider the invariance of the action (\ref{mact2}) under the infinitesimal (bosonic) transformations
 \bea
 \delta\phi^\mu= a L^\mu{}_{\nu\rho} D\phi^\nu D\phi^\rho~.
 \label{ltrans}
 \eea
 These transformations leave the action invariant provided \cite{gpt} that
 \bea
 L_{\mu\nu\rho}=L_{[\mu\nu\rho]}~,~~~\hat\nabla_{\mu} L_{\nu_1\nu_2\nu_3}=\hat\nabla_{[\mu} L_{\nu_1\nu_2\nu_3]}~,~~~
di_L c+4i_L dc=0~.
\label{lcon}
\eea
Such transformations will be useful to explore (\ref{ltcfh}).

To identify the symmetries of a particle system with action (\ref{mact2}) propagating in the supersymmetric $D=5$, ${\cal N}=1$ supergravity background, one has to match the conditions of the TCFH (\ref{tcfh2}) with those of the invariance (\ref{psusy2}) of the particle system (\ref{mact2}). For this let us consider three independent
transformations (\ref{itran2}) generated by the tensors $I^r$, $r=1,2,3$ and identify $I^r$ with the 2-form bilinears $\omega^r$ of TCFH, i.e. $I^r=\omega^r$.
Comparing the TCFH connection on $\omega^r$ with that on $I^r$ in (\ref{psusy2}), one concludes that the coupling $c$ of the particle system should be chosen as
\bea
c=2\sqrt 3\, {}^*F~.
\label{tor}
\eea
Then consistency of (\ref{tcfh2}) with  (\ref{psusy2}) after this identification   requires   that
\bea
{}^*F_{\rho\alpha\beta} \omega^{r\alpha\beta}=0~,~~~di_{\omega^r} {}^*F-3 i_{\omega^r} d{}^*F=0~.
\label{cons2}
\eea
These two conditions impose strong restrictions on the possible backgrounds for which the particle system (\ref{mact2}) admits (\ref{itran2}) as a symmetry.

Before we turn to investigate (\ref{cons2}) for various backgrounds, observe that $K$ is a Killing vector that leaves $F$ invariant.  As a result
$\delta\phi^\mu=a K^\mu$
is a symmetry of  (\ref{mact2}).

To find the backgrounds that satisfy (\ref{cons2}), let us begin with the supersymmetric backgrounds of $D=5$, ${\cal N}=1$ supergravity that admit a time-like Killing spinor, i.e. a Killing spinor such that vector bilinear $K$ is time-like \cite{gutowski}.  For such backgrounds the Killing spinor can be written as
$\epsilon=1 V$ in the conventions of \cite{review}, where $V$ is a spacetime function and the metric and 2-form flux are given as
\bea
&&ds^2=-V^4 (dt+\beta)^2+ V^{-2}  d \mathring s^2~,
\cr
&& F={\sqrt3 \over 2} d e^0-{1\over3} (d\beta)_{\rm asd}~,
\eea
with $K=\partial_t$ and $e^0=V^2 (dt+\beta)$, where $(d\beta)_{\rm asd}$ is the anti-self dual component of $d\beta$ and $d \mathring s^2$ is a 4-dimensional hyper-K\"ahler metric. In our conventions $\omega^r$ are self-dual and in addition
 $d\omega^r=0$, $i_K \omega^r=0$ and
\bea
\omega^r_{\rho\mu} \omega^{s \rho}{}_{\nu}= \delta^{rs} (V^4 g_{\mu\nu} +K_\mu K_\nu)+ \epsilon^{rs}{}_q V^2 \omega^q_{\mu\nu}~.
\label{auq}
\eea
Also one finds that ${\cal L}_K\omega^r=0$.

The first condition in (\ref{cons2}) implies that
\bea
V={\rm const}~,~~~(d\beta)_{ij} \omega^{rij}=0~.
\eea
As $V$ is constant set for convenience $V=1$. Furthermore, the equations of motion imply that $(d\beta)_{\rm asd}=0$.  As $d\omega^r=0$, set
\bea
d\beta= u_r \omega^r~,
\eea
where $u$ is a constant vector.  Without loss of generality pick $(u_r)=(1,0,0)$.  Implementing all the restrictions mentioned above, the resulting solution is expressed as
\bea
ds^2=- (dt+\beta)^2+   d\mathring s^2~,~~~
 F={\sqrt3 \over 2} \omega^1~.
\label{sol2}
\eea
The solution can be viewed locally as a circle fibration over a 4-dimensional hyper-K\"ahler manifold whose fibre $U(1)$ curvature is given by $\omega^1$.
It turns out that the last condition in (\ref{cons2}) is also satisfied for  the transformations  (\ref{itran2})  generated by $\omega^2$ and $\omega^3$.
Thus the action (\ref{mact2}) with couplings given in (\ref{sol2}) is invariant under the transformations (\ref{itran2}) generated by $\omega^2$ and $\omega^3$.

Note that this is unlike what has been encountered before in the context of supersymmetric sigma models where two supersymmetries of the type (\ref{itran2})  generated by $I^2$ and $I^3$, respectively, always  imply the existence of a third supersymmetry generated by $I^2 I^3$. However to derive this,  there have been some assumptions.  In particular  it had been taken that $I^2$ and $I^3$ are invertible and the sigma model manifold is (almost) hyper-complex. However here $\omega^2$ and $\omega^3$ are not invertible as spacetime tensors and $\omega^1$ is singled out as the curvature of the $U(1)$ bundle over the hyper-K\"ahler manifold.

It remains to solve (\ref{cons2}) for $D=5$, ${\cal N}=1$ supergravity backgrounds that admit a null Killing spinor. In such a case the Killing vector bilinear  $K=\partial_u$ is null and one can show that there is a co-frame
\bea
e^+=du+ Vdv+n_I dx^I~,~~~e^-=h^{-1} dv~,~~~e^i=h \delta^i_I dx^I~,~~~i=1,2,3~,
\eea
 where  $(u,v, x^I)$, $I=1,2,3$ are the spacetime coordinates and $V, h, n_I$ depend only on $x^I$ and $v$. Moreover one has   that
\bea
\omega^r=e^-\wedge e^r~,
\eea
and
\bea
&&ds^2=2e^- e^++ \delta_{ij} e^i e^j~,
\cr
&&F=-{1\over 4\sqrt3} \mathring \epsilon_I{}^{JK} h^{-2} (dn)_{JK} dv\wedge dx^I-{\sqrt3\over4} \mathring \epsilon_{IJ}{}^K \partial_K h dx^I\wedge dx^J~,
\eea
where  $\mathring \epsilon$ is the Levi-Civita tensor of the flat metric.

The first condition in (\ref{cons2}) for all $\omega^r$ implies that $h$ must depend only on $v$, $h=h(v)$. It turns out that this condition is also sufficient for the second condition in (\ref{cons2}) to be satisfied. There are many solutions with $h=h(v)$. For example one can take $n=0$, $h=1$ in which case the field equations imply that $V$ is harmonic function on $\bR^3$ with delta function sources and the solution a multi pp-wave. Another solution is to take $h=1$, $n=n(x^I)$.  Then  the field equations imply, see e.g. \cite{review},   that
\bea
\partial^I dn_{IJ}=0~,~~~\partial^I\partial_J V={1\over6} dn^{IJ} dn_{IJ}~,
\eea
i.e. $dn$ satisfies the Maxwell equation on $\bR^3$ and $V$ the Laplace equation with a source term. For a solution  set $n_I=\lambda_{IJ} x^J$, $\lambda$ constant 2-form and $V=(1/3) \delta^{IJ} \lambda_{IK} \lambda_{JL} x^K x^L+ V_0$, where $V_0$ is a harmonic function on $\bR^3$ with delta-function sources.

The commutators of two (\ref{itran2}) transformations can easily be computed and involve the Nijenhuis tensor of the $I$'s that generate the transformations. In the examples explored above, these do not satisfy the standard supersymmetry algebra in one dimension.  This can be easily seen as $\omega^r$ does not satisfy the algebra of imaginary unit quaternions, see (\ref{auq}). Nevertheless the commutator is a symmetry of the action (\ref{mact2}).

Next let us consider whether the 3-form bilinears (\ref{lform}) generate symmetries for the action  (\ref{mact2}). For this consider three transformation as in
(\ref{ltrans}) generated by the tenasors $L^r$ and identify $L^r=\lambda^r$.  Then consistency of (\ref{ltcfh}) with (\ref{lcon}) requires that
\bea
c=2\sqrt3 {}^*F~,~~~di_{\lambda^r} {}^*F+4i_{\lambda^r} d{}^*F=0~,~~~F_{\gamma \mu} \omega^{r\gamma}{}_\nu-F_{\gamma \nu} \omega^{r\gamma}{}_\mu=0~.
\label{conlx}
\eea
The third condition above arises from the requirement that the last term in the TCFH (\ref{ltcfh}) must vanish.

There are solutions to the conditions (\ref{conlx}) for supersymmetric backgrounds with both a timelike and null Killing spinors.  In the former case, the last condition in (\ref{conlx}) together with the field equations  imply that $V=1$ and $ (d\beta)_{\rm ads}=0$. Without loss of generality one again can choose $d\beta=\omega^1$. The spinning particles described by the  action  (\ref{mact2}) on such such backgrounds  are invariant under the transformation generated by $\lambda^1$ but they are not invariant under the transformations generated by $\lambda^2$ and $\lambda^3$.
For backgrounds with a null Killing spinor, one again finds as a consequence of the last equation in (\ref{conlx}) that $h=h(v)$. Then the analysis proceeds as for the symmetries generated by $\omega^r$ giving the same backgrounds as solutions.

The 2-form bilinears of both $D=4$ and $D=5$ supergravities that generate symmetries in the spinning particle actions  we have investigated  are not principal. This means that they do not have 2 independent eigenvalues. The existence of a principal CCKY form on a background implies the separability of the geodesic equations and some of the classical field equations,  see e.g. \cite{revky, frolov}.  Therefore one should not expect that the backgrounds we have investigated exhibit similar separability properties unless they admit additional symmetries, e.g. additional rotation or axial symmetries. To give an explicit example consider the solution of $D=5$ supergravity which is locally a circle bundle over a 4-dimensional hyper-K\"ahler manifold. We have found that a particle system in such a background admits additional fermionic symmetries. However if one chooses as a hyper-K\"ahler manifold one without additional isometries, e.g $K_3$, one should not expect that the geodesic equations of the 5-dimensional solution to be separable.

The separability properties of $D=5$ $N=1$ supergravity backgrounds    have been investigated before in \cite{Kubiznak, houri, warnick2}.  These authors explored the properties  of  the generalized   CKY equation which is the CKY equation with respect to a connection with skew-symmetric torsion, like $\hat\nabla$ in (\ref{skewcon}). In particular they considered generalized closed CKY 2-forms, i.e. 2-forms which are closed with respect to $\hat d$ the exterior derivative associated to $\hat \nabla$. The 2-form bilinears $\omega^r$ we have considered here do not satisfy the same conditions as the generalized closed CKY forms.
In particular,  $\omega^r$ satisfy the generalized CKY equation as a consequence of (\ref{tcfh2})  with skew-symmetric torsion $c$  given in (\ref{tor}).  However for general supersymmetric solutions $\omega^r$ do not satisfy the closure (or indeed the co-closure) condition with respect to $\hat\nabla$, i.e.  $\hat d \omega^r\not=0$.  Of course as a consequence of the TCFH in (\ref{tcfh2}) $\omega^r$ are closed, $d\omega^r=0$, in the standard sense. Therefore the  gravitational backgrounds investigated in \cite{Kubiznak, houri} and  in this paper are different.

\end{document}